\documentclass[
reprint,
amsmath,
amssymb,
aps,
prl,superscriptaddress,nobibnotes
]{revtex4-2} 

\usepackage{graphicx}
\usepackage{textgreek}
\usepackage{dcolumn}
\usepackage{bm}
\usepackage{siunitx}
\usepackage{braket}
\usepackage[dvipsnames]{xcolor}
\usepackage[normalem]{ulem}
\usepackage[T1]{fontenc}
\usepackage[utf8]{inputenc}
\usepackage{lmodern}
\usepackage{xargs}
\usepackage{placeins}

\setcitestyle{super}

\usepackage{hyperref}
\hypersetup{
    colorlinks,
    linkcolor={black},
    citecolor={blue!80!black},
    urlcolor={blue!80!black}
}

\begin{document}

\preprint{ACS/123-QED}

\title{Limits of Stable Near-Field Probing in Nanophotonic Traps}

\author{Johannes Piotrowski}
\thanks{equal contribution}
\affiliation{Department of Physics, Humboldt-Universit\"{a}t zu Berlin, 10099 Berlin, Germany}
\author{Constanze Bach}
\thanks{equal contribution}
\affiliation{Department of Physics, Humboldt-Universit\"{a}t zu Berlin, 10099 Berlin, Germany}
\author{Nicol\'{a}s Vera Paz}
\affiliation{Departamento de F\'{i}sica, Universidad de Concepci\'{o}n, Concepci\'{o}n, Chile}
\author{Philipp Schneeweiss}
\affiliation{Department of Physics, Humboldt-Universit\"{a}t zu Berlin, 10099 Berlin, Germany}
\author{Arno Rauschenbeutel}
\affiliation{Department of Physics, Humboldt-Universit\"{a}t zu Berlin, 10099 Berlin, Germany}

\begin{abstract}
    \looseness=-1
Near-fields around nanophotonic structures and waveguides can be used to optically interface particles ranging from atoms and molecules to microscopic biological and synthetic particles.
Due to the strong, non-linear dependence of the near-field coupling strength on the particles' position, a change of the spread of the particles' position will change their mean coupling strength.
When the particles are trapped, this position spread depends on their temperature, generally leading to temperature-dependent coupling.
Here, we experimentally demonstrate that this effect renders optical probing of trapped particles with near fields an inherently transient process.
Specifically, we trap cold atoms in a two-color dipole trap surrounding an optical nanofiber and probe them with the evanescent field of guided, resonant light. 
The scattering of this probe light heats up the atoms, leading to a decrease of the coupling strength as well as loss of atoms. We observe both effects via a concurrent decrease of the absorption signal. 
In addition, we demonstrate that the coupling strength can be recovered by cooling the atoms back to their initial temperature.
Our findings are relevant for numerous situations where stable coupling of trapped particles to a nanophotonic structure is required.
\end{abstract}

\maketitle

Nanophotonic structures and waveguides have emerged as versatile tools for optically interfacing microscopic biological and synthetic particles~\cite{Taitt2016biosensor, Toropov2021bio,Fish2022TIRF,PraveenKamath2023reviewtrapping,Wu2023TIR}, molecules~\cite{Acuna2014molecule,Arroyo2016molecule}, and laser-cooled atoms.\cite{Rychtarik2004surfacetrap,sague2007atomfiber,Aoki2006microresonator,Hakuta2012atoms,Zhou2024microring,Bouscal2024photoniccrystal} 
These interfaces are based on the particles' interaction with the near-field of the structure's photonic modes.
The latter exhibits a short intensity decay length, resulting in a strong, non-linear dependence of the interaction strength on the distance between the particles and the nanostructure.\cite{ZHANG2023review,lee2023waveguides,Nayak2018reviewnanofiber,Anetsberger2009nearfield} 
To limit the resulting motion-induced variation of the interaction strength, particles can be optically trapped close to the structure. 
However, the particles' position spread in the trap will still depend on their temperature, thereby rendering the position-averaged coupling strength in general temperature-dependent.
Moreover, when probing the trapped particles with light, the inevitable scattering of this probe light heats the particle motion.
Together, these effects preclude stable optical probing of particles near nanophotonic structures even before heating-induced particle loss occurs.\\
We observe this phenomenon with trapped and laser-cooled atoms exploring the evanescent fields of guided light near the surface of a nanofiber.\cite{vetsch2010optical,Goban2012trap} 
This system has been proven to be a highly controlled experimental platform, e.g., for studying chiral interactions~\cite{lodahl2017chiral,Suarez2025chiral}, collective emission~\cite{corzo2016Bragg,sorensen2016backscattering,Solano2017superradiance,Han2021beats,liedl2024observation,Bach2026second}, and implementing quantum memories.\cite{Gouraud2015memory,Sayrin2015storage,Corzo2019single}
At the same time, atoms trapped close to a nanofiber surface are subject to heating when scattering near-resonant light~\cite{Ostfeldt2017evanescent,Markussen2020motion,Karanikolaou2024statedependent}, in addition to a passive heating due to vibrations of the nanofiber.\cite{Reitz2013coherence,hummer2019heating} 
This heating hampers the observation of phenomena that require stable coupling between the atomic ensemble and the guided light and will ultimately eject atoms from the trap.\cite{Dawkins2011interface,Solano2017polarimetry}\\
Specifically, we measure the transient transmission of resonant probe light through an ensemble of nanofiber-coupled atoms.
We identify two timescales, one attributed to the heating-induced reduction of the coupling strength and the other to the loss of atoms. 
By laser-cooling the remaining atoms back to their initial temperature~\cite{Meng2018}, we can restore their coupling strength.
A model based on atoms in a Morse-like potential~\cite{Morse1929} allows us to match our data and estimate the heating rate applied by the probe light, as well as the cooling rate.

\section{Interfacing emitters in asymmetric potentials with an evanescent field}
\begin{figure}
    \centering
    \includegraphics[width=0.9\columnwidth]{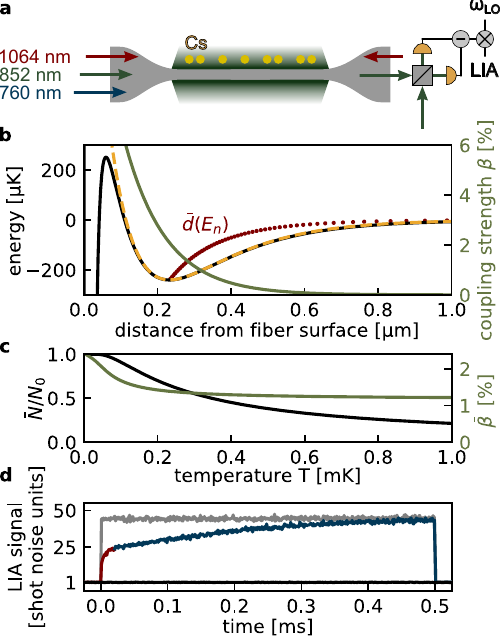}
    \caption{\textbf{Setup for lock-in amplified transmission measurements of atoms trapped and probed by evanescent fields.}
    \textbf{(a)}~Two guided light fields, red (wavelength \SI{1064}{\nano\meter}, standing wave) and blue (\SI{760}{\nano\meter}, travelling wave) detuned with respect to the atomic resonance, trap cesium atoms close to the surface the nanofiber waist of a tapered optical fiber.
    Resonant (\SI{852}{\nano\meter}) probe light is transmitted through the nanofiber and interferes with a local oscillator (LO) on a heterodyne detector. 
    The signal is lock-in amplified (LIA) at the LO frequency $\omega_\mathrm{LO}$.
    \textbf{(b)}~The evanescent fields of the red and blue detuned lasers create a trapping potential (solid black) that is well approximated by a Morse-like potential (dashed yellow). 
    This asymmetric potential causes the mean distance $\bar{d}$ (red points) of a trapped atom from the fiber surface to increase for higher motional states with quantum number $n$ and Energy $E_n$.
    The coupling strength $\beta$ (green) of atoms to the evanescent probe field decays exponentially for larger distances.
    \textbf{(c)}~Both the \textit{average} coupling strength $\bar{\beta}$ and the remaining relative atom number $\bar{N}/N_0$ decrease with temperature $T$.
    \textbf{(d)}~The transmission through the nanofiber is calculated from three LIA signals: Shot noise (black) of only the LO, reference transmission (gray) of a pulse of probe light without trapped atoms, and the signal, which exhibits a sharp initial flank (red) followed by a gradual increase (blue).
    }
\label{fig:setup}
\end{figure}
We interface emitters in an anharmonic potential, that is asymmetric with respect to its minimum, with an evanescent field by trapping cesium atoms around the nanofiber waist of a tapered optical fiber with \SI{250}{\nano\meter} radius, as sketched in Fig.~\ref{fig:setup}~a.
Two laser fields, a travelling wave, which is blue-detuned with respect to the D-line transitions (wavelength \SI{760}{\nano\meter}, power \SI{20.5}{\milli\watt}), and a red-detuned standing wave (wavelength \SI{1064}{\nano\meter}, powers \SI{1.3}{\milli\watt} and \SI{1.1}{\milli\watt}) are coupled into the tapered optical fiber.
Their evanescent fields create two diametral arrays of trapping sites along the nanofiber~\cite{vetsch2010optical}, which we load with at most one atom per site~\cite{Schlosser2002} from a magneto-optical trap through polarization-gradient molasses cooling.
The atoms on one side of the nanofiber are brought close to their motional ground state, and thus positioned close to the minimum of their trapping potential, by side-selective degenerate Raman cooling (DRC) with an additional fiber-guided laser field.\cite{Meng2018} 
The atoms on the other side are ejected during this process.\\
We calculate the radial profile of the trapping potential~\cite{lekien2004trap} from the above system parameters and draw it in Fig.~\ref{fig:setup}~b.
For distances $d\gtrsim\SI{100}{\nano\meter}$ from the fiber surface, the trap profile is well approximated by a Morse-like potential~\cite{Morse1929}
\begin{equation*}
    U = D\left[\mathrm{e}^{-2a(d-d_0)}-2\mathrm{e}^{-a(d-d_0)}\right]\,,
\end{equation*}
with a depth $D=\SI{240}{\micro\kelvin}$, stiffness $a = \SI{5.9}{\per\micro\meter}$ and its potential minimum at a distance of $d_0 = \SI{231}{\nano\meter}$, as shown in the dashed yellow line.
In this asymmetric potential, the average distance $\bar{d}$ of a trapped atom depends on its motional state with quantum number $n$ and energy $E_n$~\cite{Lima2005Morse}, see the red points in Fig.~\ref{fig:setup}~b.
The light scattered by the atoms couples to the fiber guided mode in forward direction with a strength $\beta(d)$ (green line), which decreases exponentially as $d$ increases.
From the above parameters, we calculate the temperature-dependent average coupling strength $\bar{\beta}(T)$ and average number $\bar{N}(T)$ of remaining trapped atoms with respect to the number of initially loaded atoms, $N_0$ (see Supplementary).
The results are shown in Fig.~\ref{fig:setup}~c.
We note that $\bar{\beta}(T)$ also takes the spread of distances of atoms from the fiber surface into account.
We use temperature as a parameter to discuss ensemble averages in the following but note that our independently trapped atoms do not thermalize with each other.\\
To facilitate probing, guided laser light is tuned to resonance with the D2-cycling transition of atoms in the potential minimum (wavelength \SI{852}{\nano\meter}).
The probe light provides a way to interrogate the optical depth (OD) of the trapped atoms, but its scattering also leads to their heating.
We quote the probe power in the nanofiber as $P_\mathrm{in}^\mathrm{norm}$, which is normalized to the saturation power $P_\mathrm{sat}$ taken from absorption measurements (see Supplementary). 
Probe pulses are generated from the continuous-wave output of our probe laser using acousto-optic modulators with rise times shorter than our sampling time of \SI{1}{\micro\second}.
We send these probe pulses through the nanofiber and interfere its output with a local oscillator (LO) on balanced photodiodes in a heterodyne scheme. 
The heterodyne signal is lock-in amplified (LIA) at the LO frequency $\omega_\mathrm{LO}/2\pi = \SI{5}{\mega\hertz}$, and we average data over 20 pulses.
An example LIA signal is shown in Fig.~\ref{fig:setup}~d.
The shot noise background of the LO, and the transmission signal of a probe pulse without trapped atoms, are shown in black and gray, respectively.
They are used to calculate the normalized transmission from signal with trapped atoms (colored lines).
We observe initial and long-time transmission dynamics, marked in red and blue, respectively.
Their characteristics relate to the interplay between temperature-dependent coupling $\bar{\beta}(T)$ as well as heating-related loss of trapped atoms $\bar{N}(T)$, 
We will discuss both effects in detail in the following sections.

\section{Transient transmission}
\begin{figure}
    \centering
    \includegraphics[]{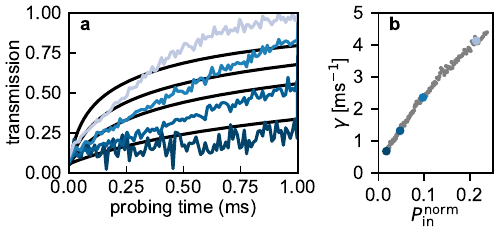}
    \caption{\textbf{Transient transmission dynamics.}
    \textbf{(a)}~As atoms are heated up and lost from their traps during the probe pulse, the transmission (colored lines) rises towards unity. 
    Black lines are simulation results for the measured normalized probe power $P_\mathrm{in}^\mathrm{norm}=0.01,\,0.05,\,0.10,\,0.22$, from bottom to top, respectively.
    \textbf{(b)}~The OD decay constant $\gamma$ depends on the probe power. Colored data points correspond to the traces in (a).
    }
    \label{fig:slow_heating}
\end{figure}
Figure~\ref{fig:slow_heating}~a shows measured transients in the transmission of \SI{1}{\milli\second} long probe pulses.
For $P_\mathrm{in}^\mathrm{norm} \ll 1$ and small OD, the transmission curves are well approximated by a double exponential dependence
\begin{equation}\label{eq1}
    \mathrm{transmission} = \left[1-2\bar{\beta}(t)\right]^{2\bar{N}(t)} \approx \mathrm{e}^{-{O\!D}_0\mathrm{e}^{-\gamma t}}\,,
\end{equation}
with an initial OD of ${O\!D}_0$ and an OD decay constant $\gamma$. 
Here, the resonant amplitude transmission coefficient of a single atom is $1-2\beta$.
First, we extract $\gamma$ from fitting Eq.~\eqref{eq1} to the transmission data in Fig.~\ref{fig:slow_heating}~a (fits not shown). 
The results are shown in Fig.~\ref{fig:slow_heating}~b, where $\gamma$ increases with $P_\mathrm{in}^\mathrm{norm}$.
We ensure that the probing time is much shorter than the passive lifetime of atoms in their traps $\tau_0 = \SI[separate-uncertainty=true]{84(4)}{\milli\second}$ (see Supplementary Fig.~\ref{fig:lifetime}).
This observation indicates that the probe light is the dominant heating mechanism for our trapped atoms.\\
For a deeper understanding of the heating dynamics in an evanescent field, we simulate the time dependence of $\bar{N}$ and $\bar{\beta}$ in Eq.~\eqref{eq1} considering their temperature dependence in Fig.~\ref{fig:setup}~c.
We assume that the temperature rises from an initial value of $T_0\approx \SI{1}{\micro\kelvin}$~\cite{Meng2018} after ground state preparation according to the heating rate 
\begin{equation*}
    \mathrm{d}T/\mathrm{d}t = R_\mathrm{sc}(s,T) \Delta T(T)\,.
\end{equation*}
The scattering rate $R_\mathrm{sc}$, and the heating per scattering $\Delta T$, are themselves dependent on the saturation parameter $s$ of the atom and the temperature $T$ (for details see Supplementary).
In our trap configuration, $\mathrm{d}T/\mathrm{d}t$ is also higher than expected for simple recoil ($\approx\SI{1}{\kelvin\per\second}$ for $s=1$) due to dipole-force fluctuation heating caused by state-dependent potentials.\cite{lekien2013statedependent,MartinezDorantes2018statedependent,Karanikolaou2024statedependent} 
Typical passive heating rates in our system, reported as $0.34\,\mathrm{phonons/ms}$~\cite{Albrecht2016gradients} or $\SI{3}{\milli\kelvin\per\second}$~\cite{Reitz2013coherence}, would already be exceeded by $\mathrm{d}T/\mathrm{d}t$ for probe powers on the \SI{10}{\femto\watt} scale, or $s \sim 0.0001$.
In order to obtain sufficient signal, we used much higher probe powers in the $\SI{10}{\pico\watt}$ scale.
This consequently reduces the lifetime of atoms in our traps below $\tau_0$, limiting the available probing time.\\
The simulation results (black lines) rely solely on system parameters with no fitting and reproduce the measured transmissions in Fig.~\ref{fig:slow_heating}~a reasonably well. 
The calculated value $\bar{\beta}(T\rightarrow\infty)=0.012$ matches independent absorption measurements, from which we also extract $N_0 = 29$ (see Supplementary).

\section{Early-time heating effect}
\begin{figure}
    \centering
    \includegraphics{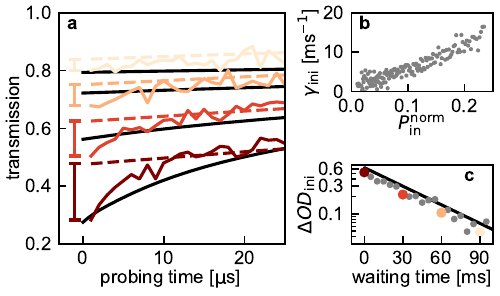}
    \caption{\textbf{Initial transmission flank.}
    \textbf{(a)}~The transmission at the beginning of the probe pulse is lower (marked by vertical bars) than extrapolated from the dynamics at longer timescales (dashed lines). 
    From bottom to top, we add 0, 30, 60 and \SI{90}{\milli\second} waiting time, respectively, between the ground-state preparation of the atoms and the probing pulse ($P_\mathrm{in}^\mathrm{norm}=0.27$). 
    For longer waiting times, the initial flank vanishes in agreement with the simulations (black lines). 
    \textbf{(b)}~The initial OD decay constant $\gamma_{\mathrm{ini}}$ increases with probe power and is typically a factor of 3 larger than the long-time OD decay constant $\gamma$.
    \textbf{(c)}~The initial drop in OD ($\Delta {O\!D}_{\mathrm{ini}}$), as calculated from the measured transmission increase (data points) and simulated (line) from the dynamics of $\bar{\beta}$, decreases for longer waiting times.
    Colored data points correspond to the traces in (a).
    }
    \label{fig:hotflank}
\end{figure}
The initial part of the measured transmission curves deviates from the double exponential behavior predicted by the right-hand side of Eq.~\eqref{eq1}.
Figure~\ref{fig:hotflank}~a shows double exponentials (dashed lines) fitted to the transmission data from $10$ to $\SI{500}{\micro\second}$.
These fits overestimate ${O\!D}_0$ at the beginning of the probe pulse by an amount $\Delta {O\!D}_{\mathrm{ini}}$ (vertical bars) because of a steeper transmission increase emerging in the first microseconds of probing.
The OD decay constants $\gamma_{\mathrm{ini}}$ fitted to these initial flanks up to \SI{10}{\micro\second} increase with rising input power $P_\mathrm{in}^\mathrm{norm}$, see Fig.~\ref{fig:hotflank}~b.
We further observe that $\Delta {O\!D}_{\mathrm{ini}}$ decreases, when we introduce a waiting time between the initial ground state preparation of the atoms and the probing pulse, see Fig.~\ref{fig:hotflank}~c.\\ 
Simulations based on our model (black lines in Fig.~\ref{fig:hotflank}~a) reproduce the measured initial transmission flank.
For the short timescale considered here, the dynamics is dominated by the fast change of $\bar{\beta}$, leading to the steep initial flank with $\gamma_{\mathrm{ini}}$.
For longer wait times, $\bar{\beta}(T_0)$ approaches $\bar{\beta}(T\rightarrow\infty)$, the initial flank vanishes, and the dynamics follows the long-time behavior studied in Fig.~\ref{fig:slow_heating}, which is dominated by atom loss.
For simulating the transmission after a certain waiting time, we assume $T_0$ to increase according to a passive heating rate of $\SI{6}{\milli\kelvin\per\second}$ (twice the value reported in~\cite{Reitz2013coherence}).
With this heating rate, the modelled $\Delta {O\!D}_{\mathrm{ini}}$ (black line in Fig.~\ref{fig:hotflank}~c) matches the experimental values.

\section{Recovering coupling strength by cooling}
\begin{figure}
    \centering
    \includegraphics{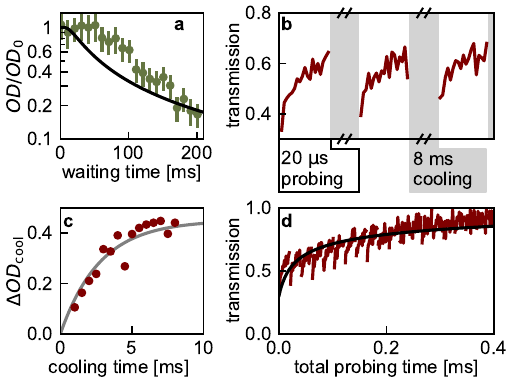}
    \caption{\textbf{Partial recovery of transient OD loss.}
    \textbf{(a)}~When applying \SI{20}{\milli\second} of DRC before the OD measurement, the OD loss during waiting time is driven by atom loss and shows a plateau in both the data (points with error bars of one standard deviation of the fit) and simulation (line).
    \textbf{(b)}~\SI{20}{\micro\second} probe pulses are interleaved with \SI{8}{\milli\second} of DRC, partially recovering the OD. 
    \textbf{(d)}~The combined transmission curve (red) of 20 probe pulses, with the cooling periods removed, matches the long-time dynamics of the simulation (black) of Eq.~\eqref{eq1}.
    This implies that permanent atom loss from the traps still occurs during probing while only the remaining atoms can be cooled during the cooling cycles.
    \textbf{(c)}~We quantify cooling performance from the OD increase $\Delta {O\!D}_\mathrm{cool}$ during the cooling step.
    $\Delta {O\!D}_\mathrm{cool}$ (red points) rises with longer cooling time, and an exponential fit (gray line) returns a time constant of \SI{360}{\per\second}.
    }
    \label{fig:recooling}
\end{figure}
We can partially reverse the initial OD loss by again cooling the atoms close to their motional ground states via DRC, thereby recovering $\bar{\beta}(T\approx\SI{1}{\micro\kelvin})$. 
This observation underpins the important insight that a loss of OD is not necessarily caused by a loss of atoms. 
Moreover, by cooling the atoms, we can separately observe the dynamics of $\bar{\beta}$ and $\bar{N}$. \\
Usually, the OD decay of an atomic ensemble is well described by an exponential function, where the decay time is given by the passive atomic lifetime $\tau_0$.
In Fig.~\ref{fig:recooling}~a, we apply \SI{20}{\milli\second} of DRC right before measuring the OD. 
In this case, the remaining OD after DRC (green points) remains roughly constant for about \SI{50}{\milli\second}.
For the simulation (black line), we assume $\bar{\beta}(T) = \bar{\beta}(T_0=\SI{1}{\micro\kelvin})$ and find a similar, albeit shorter, plateau.
Given that the temperature after DRC is close to $T_0$ for all data points shown, the observed loss of OD is mainly caused by irreversible loss of atoms.
This loss occurs during the wait time, where the temperature rises up to $T_\mathrm{wait}$, at which only a number $\bar{N}(T_\mathrm{wait})$ of atoms remains trapped and can be cooled again.
The plateau of the OD in Fig.~\ref{fig:recooling}~a hence has the same origin as the plateau of $\bar{N}(T)$ in Fig.~\ref{fig:setup}~c, namely that atom loss is minimal for temperatures much lower than the trap depth $D=\SI{240}{\micro\kelvin}$.\\
Fig.~\ref{fig:recooling}~b unveils the contribution of $\bar{\beta}(T)$ to the OD. 
Here, we interleave heating and cooling cycles, during \SI{20}{\micro\second} probing pulses with $P_\mathrm{in}^\mathrm{norm} = 0.26$, and \SI{8}{\milli\second} of applying DRC, respectively.
The total time elapsed during one experimental run ($20\times\SI{8.02}{\milli\second}=\SI{160.4}{\milli\second}$) is shorter than the lifetime under DRC, which exceeds \SI{1}{\second} (see Supplementary Fig.~\ref{fig:lifetime}).
Each probing pulse creates a flank in transmission with $\gamma_{\mathrm{ini}}\approx \SI{20}{\per\milli\second}$, while the total probe trace with the cooling times removed shows an overall double exponential transmission increase with $\gamma\approx \SI{6}{\per\milli\second}$, see Fig.~\ref{fig:recooling}~d.
Both decay constants agree with the expectations from Fig.~\ref{fig:slow_heating}~b and Fig.~\ref{fig:hotflank}~b, respectively.
According to our model, after \SI{20}{\micro\second} of probing, the atoms are heated to about \SI{100}{\micro\kelvin}.
At this temperature, less than $10\%$ of atoms are lost but $\bar{\beta}$ has already decreased to about $70\%$ of its original value.
Thus, the observed initial flanks are mainly due to the temperature-induced loss in coupling strength.\\
We estimate the cooling rate in the absence of probe light via the amount $\Delta {O\!D}_\mathrm{cool}$ of recovered OD during the cooling cycles.
In Fig.~\ref{fig:recooling}~c, we vary the duration of the cooling cycles and observe that $\Delta {O\!D}_\mathrm{cool}$ exponentially approaches a maximum value of about 0.4 with a time constant of \SI{360}{\per\second}.
According to our model, this maximum value corresponds to a change of $\bar{\beta}$ from $0.017$ to $0.024$ as we cool the atoms from \SI{100}{\micro\kelvin} to \SI{1}{\micro\kelvin}, respectively, according to a cooling rate of about \SI{100}{\milli\kelvin\per\second}.

\section{Discussion}
We investigated the transient transmission of light through atoms that are optically trapped and probed with the evanescent field surrounding an optical nanofiber. 
We identified two contributions to the observed decay of optical depth caused by the probing-related heating of the atoms:
The short-time dynamics is caused by a temperature-related change of the spread and mean position of  the atoms. 
In conjunction with the distance-dependence of the coupling strength, this results in a reduction of mean coupling strength and, hence, optical depth of the atomic ensemble.
This effect is smaller for atoms with higher initial temperature and can be partially reversed by cooling the atoms.
The remaining loss of optical depth after each probing period is due to the heating-related loss of atoms, an effect that dominates the long-time dynamics.
The combination of these two contributions results in richer dynamics of the optical depth than the commonly assumed exponential decay.\\
Our experimental platform realizes a model system for near-field probing of particles that are trapped in a Morse-like potential. 
However, the effects described are expected to occur in most trapping potentials and generally limit the stability of near-field probing.
In principle, this limitation could be overcome by cooling the particles with a rate that sufficiently exceeds the probing-related heating rate.
This would then enable longer measurement times and  more homogeneous coupling of the individual particles.
However, such alternative cooling schemes may be challenging and have adverse effects, like hampering atomic state preparation in our case.
Our findings are relevant for numerous situations where stable coupling of trapped particles to a nanophotonic structure is required. 
For our platform, these encompass spin squeezing~\cite{Ostfeldt2017evanescent,Beguin2018inhomogeneous} or the observation of collective radiative effects.\cite{Mahmoodian2018transport,Prasad2020correlating} 
More generally, they will also have an impact on the performance of other types of near-field-based sensors and interfaces.

\subsection{Author contributions}
C.B. and N.V.P. set up the heterodyne detection scheme. 
C.B and J.P. conducted the measurements and analyzed the data. 
P.S. and A.R. conceived the experiment and supervised the work. 
J.P., with input from all authors, performed the simulations and wrote the manuscript.
Corresponding author: J.P., {johannes.piotrowski@hu-berlin.de}

\subsection{Acknowledgments}
\begin{acknowledgments}
We thank F. Tebbenjohanns for input during the conception of this project, and R. Pennetta, L. Pache and J. Volz for discussions.
J.~P. acknowledges funding by BERLIN QUANTUM endowed by the Innovation Promotion Fund of the city of Berlin. 
N.~V. acknowledges funding from the National Agency for Research and Developement, through ANID-Subdirecci\'{o}n de Capital Humano/Doctorado Nacional/2022-21221251. 
This project was financially supported by the Alexander von Humboldt Foundation in the framework of the Alexander von Humboldt Professorship endowed by the Federal Ministry of Education and Research.
\end{acknowledgments}

\bibliography{arxiv}
\newpage
\vspace{0pt}

\setcounter{equation}{0}
\setcounter{figure}{0}
\setcounter{section}{0}
\renewcommand{\thefigure}{S\arabic{figure}}
\renewcommand{\theequation}{S\arabic{equation}}
\renewcommand{\thesection}{S\arabic{section}}

\renewcommand{\theHfigure}{S\arabic{figure}}
\renewcommand{\theHequation}{S\arabic{equation}}
\renewcommand{\theHsection}{S\arabic{section}}

\begin{center}
\textbf{\large Supplemental Materials}
\end{center}
\FloatBarrier
\section{Lifetimes in the nanofiber traps}\label{supp:lifetime}

\begin{figure}[]
    \centering
    \includegraphics[]{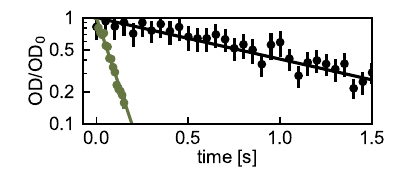}
    \caption{
    Passive lifetime (green) and lifetime under DRC (black) of trapped atoms.  Exponential fits (lines) return $\tau_\mathrm{DRC} = \SI[separate-uncertainty=true]{1114(78)}{\milli\second}$ and the passive lifetime $\tau_0 = \SI[separate-uncertainty=true]{84(4)}{\milli\second}$, respectively.
    }
\label{fig:lifetime}
\end{figure}

To measure the lifetimes of atoms in the traps, we prepare ground-state-cooled atoms in the traps around the nanofibers.\cite{Meng2018} 
We then measure the optical depth (OD) by fitting a saturated Lorentzian lineshape to a transmission spectrum around the atomic resonance~\cite{vetsch2010optical} directly after preparation with no waiting time (${O\!D}_0$) and after a waiting time (${O\!D}$).
The error bars represent one standard deviation of the fit.
In Fig.~\ref{fig:lifetime}, we show the remaining OD and fit the lifetimes without cooling (green, $\tau_0 = \SI[separate-uncertainty=true]{84(4)}{\milli\second}$) and while applying degenerate Raman cooling (DRC) during the waiting time (black, $\tau_\mathrm{DRC} = \SI[separate-uncertainty=true]{1114(78)}{\milli\second}$). 

\section{Absorption measurement}\label{supp:saturation}
\begin{figure}[]
    \centering
    \includegraphics[]{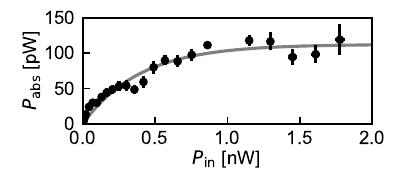}
    \caption{
        Absorbed power $P_{\mathrm{abs}}$ as a function of input power $P_{\mathrm{in}}$. 
        For large $P_{\mathrm{in}}$, $P_{\mathrm{abs}}$ saturates and reaches an asymptotic value of $P_{\mathrm{abs}}^{\max} = \SI[separate-uncertainty=true]{112(5)}{\pico\watt}$.
    }\label{fig:Psat}
\end{figure}

In Fig.~\ref{fig:Psat}, we measure the absorbed power of a laser beam through an ensemble of atoms as a function of input power.
From the exponential fit to the saturation measurement we extract the maximum absorbed power $P_{\mathrm{abs}}^{\max} = \SI[separate-uncertainty=true]{112(5)}{\pico\watt}$.
The power a single cesium atom can absorb from an excitation laser field under steady-state driving is $P_\mathrm{Cs} = \mathrm{hc}\Gamma/2\lambda = \SI{3.8}{\pico\watt}$, with $\Gamma = 2\pi\cdot\SI{5.22}{\mega\hertz}$ and $\lambda = \SI{852}{\nano\meter}$ being the full width at half maximum and the wavelength of the D2 line of cesium, respectively. 
We thus load on average $N = P_{\mathrm{abs}}^{\max} / P_\mathrm{Cs}  = \num[separate-uncertainty=true]{29(1)}$ atoms.
Together with the measured initial OD of \num[separate-uncertainty=true]{1.23(0.18)} and ${O\!D} = 4\beta N$, we can estimate our coupling strength $\beta = \num[separate-uncertainty=true]{0.011(0.002)}$ and saturation power $P_{\mathrm{sat}} = \mathrm{hc}\Gamma/8\beta\lambda = \SI[separate-uncertainty=true]{91(14)}{\pico\watt}$.
The measured $\beta$ matches the calculated $\bar{\beta}(T\rightarrow\infty)=0.012$, which is to be expected as the atoms are strongly heated during the saturation measurement.
Probe power values in the main text are normalized by $P_{\mathrm{sat}}$ and denoted $P^{\mathrm{norm}}_{\mathrm{in}}$.
We do this to distinguish them from the saturation parameter $s$, referring to the average intensity impinging on the atoms relative to the saturation intensity, which we consider in the simulations of the transient transmission behavior.

\section{Modeling transmission}\label{supp:overlap}
\begin{figure}[]
    \centering
    \includegraphics[]{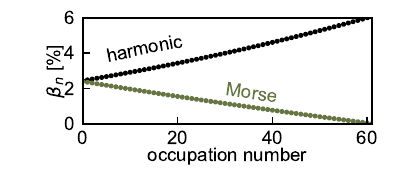}
    \caption{
        Integrated coupling strengths $\beta_n$ for the wavefunctions of a harmonic potential (black) and Morse-like potential (green) for $\Omega/2\pi = \SI{161}{\kilo\hertz}$.
    }\label{fig:betan}
\end{figure}

We model the radial trapping potential as a Morse-like potential with the parameters in the main text.
The coupling strength $\beta$ is defined as the ratio of the partial decay rate of an atom into the forward-propagating nanofiber-guided mode and the total decay rate~\cite{Scheel2015directional} (drawn in Fig.~1~b of the main text).
Our potential comes with the wavefunctions $\Psi_n$ for $n=[0,61]$ bound states with energies $E_{n} = \hbar\Omega \left(n + 0.5\right) - \hbar^{2} \Omega^{2} \left(n + 0.5\right)^{2}/4 D$ and mean distances $\bar{d_n}$ (drawn as red points in  Fig.~1~b of the main text)~\cite{Lima2005Morse}.
Here, $\Omega$ the effective radial frequency in the harmonic approximation at the trap minimum. 
The increase of the mean distance with the occupation number already shows an important difference to the harmonic potential, where the mean position remains at the center of the potential for all occupation numbers.
However, this mean picture does not capture the change in spread of the distributions.
To this end, we use the wavefunctions $\Psi_n$ to average the coupling strength $\beta$ over the radial distance and obtain
\begin{equation}\label{eq:betan}
    \beta_n = \int_0^\infty \left|\Psi_n(r)\right|^2\beta(r) dr\,.
\end{equation}
In Fig.~\ref{fig:betan}, we compare this result to the same integral when averaging over the wavefunctions of the harmonic oscillator with the same frequency $\Omega$.
It is apparent that $\beta_n$ increases for a harmonic potential, while it tends towards zero at the highest bound state $n=61$ for the Morse-like potential.
To model the average coupling strength of an ensemble of atoms, we weight $\beta_n$ with the thermal distribution in the Morse-like potential
\begin{equation*}
    P_{n} (T) = \frac{e^{- \frac{E_{n}}{k_{B}T}} \left(1 - \mathcal{H}\left(n - 61\right)\right)}{\sum_{n=0}^{61} e^{- \frac{E_{n}}{k_{B}T}}}\,,
\end{equation*}
which we truncate at the highest bound state with the Heaviside function $\mathcal{H}$.
The resulting $\bar{\beta}(T)$ is shown in Fig.~\ref{fig:setup}~c of the main text.
As this is the coupling strength for atoms that are still trapped, we model their fraction $\bar{N}(T)/N_0  = 1-e^{- \frac{D}{k_{B}T}}$, \textit{i.e.}, the part of the distribution with energy below the trap depth $D$.
We show $\bar{N}(T)/N_0$ it in the same Fig.~\ref{fig:setup}~c of the main text.\\
To simulate the transmission dynamics, we assume a heating rate $\mathrm{d}T(s,T)/\mathrm{d}t$, which is proportional to the saturation parameter $s = P_\mathrm{in}/P_\mathrm{sat} = P_\mathrm{in}\frac{8\lambda\bar{\beta}(T)}{\mathrm{hc}\Gamma}$.
Considering the dependences on saturation and temperature of the heating rate in the following section, we integrate $T(t)$ numerically and obtain the transmission as $\left[1-2\bar{\beta}(T)\right]^{2\bar{N}(T)}$. 
The input power $P_\mathrm{in}/P_\mathrm{sat}(T\rightarrow\infty) = P_\mathrm{in}^\mathrm{norm}$ is set according to the experimental values quoted in the figure legends.

\section{Heating rate}\label{supp:heatingrate}
\begin{figure}[]
    \centering
    \includegraphics[]{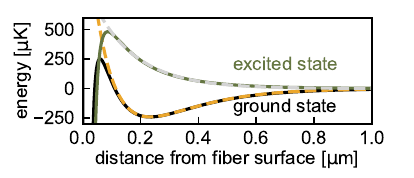}
    \caption{
        The potential for atoms in the ground (black) and excited (green) states can be approximated by a Morse-like potential (dashed yellow) and exponential repulsive potential (dashed gray), respectively.
    }\label{fig:excitedstate}
\end{figure}

\begin{figure}[]
    \centering
    \includegraphics[]{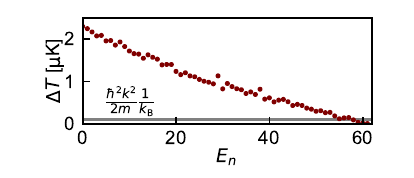}
    \caption{
        Mean temperature $\Delta T$ gained from residing in the excited state potential for an atom starting with motional energy $E_n$ in the electronic ground state. 
        The gray line is the recoil energy divided by $\mathrm{k_B}$. 
    }\label{fig:heatingrate}
\end{figure}

\begin{figure}[]
    \centering
    \includegraphics[]{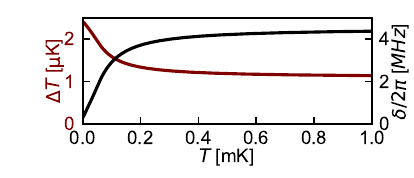}
    \caption{
    Mean temperature increase $\Delta T$ per photon scattered (red) as the sum of photon recoil and the temperature-dependent dipole force fluctuation heating in the excited state.
    The temperature-dependent detuning $\delta$ (black) is calculated from the splitting of the electronic ground and excited state potentials.
    }\label{fig:DeltaTT}
\end{figure}

A basic estimate of an upper bound to the heating rate due to recoil can be obtained by the recoil energy $\frac{\hbar^2k^2}{2m}$ of a single photon with momentum $k$ imparted on a cesium atom with mass $m$. 
Together with the scattering rate
\begin{equation}\label{eq:Rsc}
    R_\mathrm{sc} = \frac{\Gamma}{2}\frac{s}{1+s+{(2\delta/\Gamma)}^2}\,,
\end{equation}
we obtain $\mathrm{d}T/\mathrm{d}t = R_\mathrm{sc}\frac{\hbar^2k^2}{2m}\frac{1}{k_\mathrm{B}} \approx \SI{1}{\kelvin\per\second}$ at saturation $s=1$ and detuning $\delta = 0$. 
However, using this value in the simulations results in much slower increase in transmission than observed in the experiment.
Two effects related to the excited state of our two-level system influence the heating in our experiment.
As our two-color trap uses non-magic wavelengths (\SI{1064}{\nano\meter} and \SI{760}{\nano\meter}), atoms in the excited state ($\ket{6P_{3/2}, F = 5}$) experience a different potential than the ones in the ground state ($\ket{6S_{1/2}, F = 4, m_F = -4}$).\cite{lekien2013statedependent,Karanikolaou2024statedependent} 
Specifically, the potential of the excited state with the highest eigenenergy, shown in green in Fig.~\ref{fig:excitedstate}, does not trap the atoms and is purely repulsive over the extent of the bound states of the ground state. 
This renders both the heating $\Delta T$ per scattered photon and the detuning $\delta$ temperature-dependent. 
To calculate the former, we solve the atomic motion in the excited state potential for a duration sampled from an exponential distribution with the excited state lifetime of $1/\Gamma\approx\SI{30}{\nano\second}$, starting with samples of initial positions and momenta corresponding to energies $E_n$.
We note that this approach is only valid for the low saturations examined here, whereas a saturated atom will remain in its excited state for a shorter time only.
For a large number of samples ($\sim 10^5$), this method produces the mean temperature increase $\Delta T$ that we plot in Fig.~\ref{fig:heatingrate}.
This heating from a single scattering event exceeds the maximum recoil imparted by scattering a photon of $\frac{\hbar^2k^2}{2m}$ (gray line).
When the atom occupies low energy motional states, it is on average close to the potential minimum and thus experiences a strong gradient after being optically excited, leading to dipole force fluctuation heating in addition to recoil heating.
We convert this into a temperature-dependent heating $\Delta T(T) = \left[ \Delta T(E_n) + \frac{\hbar^2k^2}{2m}\frac{1}{k_\mathrm{B}} \right] P_n(T)$ and present it in Fig.~\ref{fig:DeltaTT}.\\
The distance-dependent splitting between electronic ground and excited state also leads to variable detuning.
Equivalent to Eq.~\eqref{eq:betan}, we obtain $\delta_n$ by integrating this splitting over the wavefunctions $\Psi_n$ of the Morse-like potential and convert it to $\delta(T) = \delta_n P_n(T)$, as shown in Fig.~\ref{fig:DeltaTT}.
We replace this result in Eq.~\eqref{eq:Rsc} to obtain $\mathrm{d}T/\mathrm{d}t(s,T) = R_\mathrm{sc}\left(s,\delta(T)\right)\Delta T(T)$, which is then used in the simulations throughout the main text.

\end{document}